\documentclass[%
 aip,
 jcp,%
 amsmath,amssymb,
]{revtex4-1}

\usepackage{color}
\usepackage{graphicx}
\usepackage{dcolumn}
\usepackage{bm}

\begin{document}

\title {Nature of the Effective Interaction Between Dendrimers}

\author{Taraknath Mandal}
 \email{taraknath@physics.iisc.ernet.in}

\author{Chandan Dasgupta}
 \email{cdgupta@physics.iisc.ernet.in}

\author{Prabal K Maiti}
 \email{maiti@physics.iisc.ernet.in}

\affiliation{
Centre for Condensed Matter Theory, Physics Department, Indian Institute of Science, Bangalore-560012, India.
}

\begin{abstract}
We have performed fully atomistic classical molecular dynamics (MD) simulations to calculate the effective interaction between two polyamidoamine (PAMAM) dendrimers. Using the umbrella sampling (US) technique, we have obtained the potential of mean force (PMF) between the dendrimers and investigated the effects of protonation level and dendrimer size on the PMF. Our results show that the interaction between the dendrimers can be tuned from purely repulsive to partly attractive by changing the protonation level. The PMF profiles are well-fitted by the sum of an exponential and a Gaussian function with the weight of the exponential function dominating over that of the Gaussian function. This observation is in disagreement with the results obtained in previous analytic [Macromolecules \textbf{34}, 2914 (2001)] and coarse-grained simulation [J. Chem. Phys. \textbf{120}, 7761 (2004)] studies which predicted the effective interaction to be Gaussian.
\end{abstract}

\maketitle

\section{INTRODUCTIONS}

In recent years, dendrimer molecules have attracted a significant amount of research interest because of their potential applications in the field of medicine~\cite {boas2004dendrimers, esfand2001poly, jain2010poly, jayamurugan2006synthesis, tekade2008dendrimers}, electronics ~\cite {balzani2003light, astruc2010dendrimers, ma2007functionalized, astruc2008ferrocenyl, crooks2001dendrimer}  and synthesis of nanoparticles~\cite {esumi1998preparation, balogh1998poly, zhao1998preparation, endo2005synthesis, garcia1999preparation}. These molecules are also very important in the field of nanotechnology. Self-assembly of nanoparticles mediated by dendrimers is becoming a growing field of research~\cite{srivastava2005controlled, frankamp2002controlled, daniel2003nanoscopic, ofir2008polymer, percec2004self, percec2010self, gibson2002cooperative, rosen2009dendron, zimmerman1996self, mandal2013engineering}. Dendrimers are used as spacer to control the inter-particle distance in nanocomposites, which has a strong influence on the optical~\cite {srivastava2005controlled}, electronic and magnetic properties~\cite{frankamp2005direct} of the nanocomposites. However, control over the decoration of the nanoparticles into a desired network requires a good understanding of the nature and strength of the interaction between the dendrimers. The structural~\cite{maiti2004structure, maiti2005effect, maiti2008counterion} and thermodynamic ~\cite {lin2005dynamics, maiti2009diffusion} properties of dendrimers and the interaction of dendrimers with other molecules~\cite{nandy2010dna, maingi2012pamam, nandy2012interaction, tian2010complexation, ma2013theoretical, lewis2012mean, lewis2013complexation, lewis2013interactions} have been investigated widely. 
However, there is no clear understanding of the effective interaction between dendrimers at the atomistic level. Recently, few theoretical studies have attempted to calculate the effective potential/force between dendrimers. Likos {\it et al.}~\cite{likos2001soft, likos2002gaussian} have constructed a mean-field theory based expression that describes the effective potential energy between two dendrimers. Their study predicts that the interaction between two fourth-generation (G4) dendrimers is ultra-soft and repulsive with a Gaussian shape. Subsequently, employing Monte Carlo (MC) and MD techniques, they found that the effective interaction between two dendrimers can be fitted by a sum of two Gaussian functions~\cite {gotze2004tunable}. The first function has center at the origin and the second one, centered away from this point, provides a small correction to the first function. However, in these models, the protonation level of the dendrimers and the presence of solvent and counterions, which may play important roles in the effective interaction, are not considered.  Tian {\it et al.}~\cite{ma2011coarse} have considered the effects of charges and counterions on the effective interaction between dendrimers. Using coarse-grained MD simulations, they showed that the effective interaction  depends strongly on the charges residing on the dendrimers: it can be tuned from completely repulsive to partly attractive just by changing the protonation level of the dendrimers. However, these results contradict recent results obtained by Hui\ss mann {\it et al.}~\cite {huissmann2011effective} which predict the effective interaction between the dendrimers to be always repulsive, irrespective of the charge of the dendrimers.  \\

To provide a better understanding of the nature of the effective interaction between dendrimers, which may help in resolving the controversies mentioned above, we have performed fully atomistic classical MD simulations to obtain the PMF between two PAMAM dendrimers. All earlier numerical work on this problem is based on monomer-resolved coarse-grained simulations which may not capture microscopic details such as atomic-scale fluctuations and the hydration layer around the solute. To capture all these details, we have employed all-atom classical MD simulations to explore the effects of charge (protonation level), size (generation) of the dendrimers and salt concentration in the solution on the PMF between two dendrimers. We find that the PMF can be tuned from purely repulsive to partly attractive by changing the protonation level. The PMF profiles are well-fitted by the sum of an exponential and a Gaussian function with the weight of the exponential function dominating over that of the Gaussian function. 

\subsection{METHODOLOGY}

The structures of the PAMAM dendrimers were generated using a Dendrimer Builder Toolkit~\cite{maingi2012dendrimer} developed in-house. Initially built structures were equilibrated for 20 ns and two copies of the equilibrated dendrimer were placed near each other. This entire complex was then solvated in a TIP3P~\cite{jorgensen1983comparison} water box using the xleap module of the AMBER~\cite{pearlman1995amber} package. The total number of atoms in the simulated system varied between 51639 and 151219 in different cases. Interactions between dendrimer atoms were described by GAFF~\cite{wang2004development}. Recent calculations in which this force field was used to model the dendrimers have successfully reproduced several experimental observations~\cite{maingi2012dendrimer}. Fully atomistic MD simulations were performed using the AMBER software package. Before subjecting the system to dynamical evolution, bad contacts between the water molecules and the complex solute were removed by the conjugate gradient method. This energy minimized structure was then gradually heated from 0 to 300 K. SHAKE~\cite{ryckaert1977numerical} constraint was imposed on the bonds involving hydrogen atoms which allowed us to use a relatively large time step of 2 fs. Umbrella sampling (US) techniques were used to calculate the PMF between the dendrimers. The center-to-center distance of the dendrimers was chosen as the reaction coordinate in the PMF calculation. A harmonic potential was used as the biasing potential in the US to restrain the distance between the centers of mass of the dendrimers. The interaction between the dendrimers was sampled over 50-70 equally spaced windows. In each window, the system was equilibrated for 1-2 ns and the resulting structure was used as the starting configuration for the next window. The weighted histogram analysis method was used to obtain the PMF from the biased simulation runs.

\subsection{RESULTS AND DISCUSSIONS}

Fig. 1(a) and 1(b) show the PMF between the dendrimers for the nonprotonated and protonated cases, respectively. At high pH conditions for which the dendrimers are charge neutral, the effective interaction between them is attractive for center-to-center distances exceeding a characteristic value. For distances smaller than this value, the PMF is repulsive because of the steric interactions between the dendrimer atoms. The strength of the attractive interaction, i.e. the energy released at the aggregation of two dendrimers, increases with the generation of the dendrimers. In the attractive region of the PMF, the dendrimer surfaces are in contact and interact with each other via the terminal groups ~\cite{carbone2009molecular}. So a possible reason for the stronger interaction between the higher generation dendrimers is the presence of a larger number of terminal amine groups at the dendrimer surface. In contrast to the nonprotonated case, at low pH when the terminal amine groups are fully charged (protonated), the effective interaction between the dendrimers is repulsive for all values of the center-to-center distance. To understand the origin of the attractive region in the PMF profile of nonprotonated dendrimer, we have calculated the change in dendrimer-dendrimer and dendrimer-solvent potential energy in this region. From the Fig. 2(a), we observe that the dendrimer-dendrimer potential energy decreases while the dendrimer-solvent potential energy increases in this region. However, there is a net decrease in the total potential energy. We find that mostly the van der Waals interaction contributes in the total potential energy change (inset of figure 2(a)).~\cite{footnote1}. Thus the van der Waals interaction is responsible for the attractive interaction between the nonprotonated dendrimers. Our result is consistent with the previous study~\cite{carbone2009molecular} which predicted that the van der Waals interaction and hydrogen bonding are the dominant interaction between the nonprotonated (charge neutral corresponding to high pH) dendrimers. On the other hand, in case of protonated dendrimers (corresponding to neutral pH), the dendrimer energy increases as they approach each other (figure 2(b)). This is due to the strong electrostatic repulsion between the dendrimers. The dendrimer-ion energy decreases due to the fact that larger number of ions is closer to the dendrimer units when dendrimers are close to each other. If the dendrimers are separated, the ions which are close to first dendrimer, remain at a larger distance from the second dendrimer. Thus the magnitude of the dendrimer-ion potential energy is larger when the dendrimers are closer to each other. We observe that there is a net increase in the total potential energy as the protonated dendrimers approach each other. Note that the contribution of the van der Waals interaction to the total potential energy change is relatively smaller (inset of figure 2(b)) which suggests that the electrostatic interaction has a significant contribution to the PMF between protonated dendrimers.

\vspace{0.1cm}

Our results are qualitatively consistent with those obtained by Tian {\it et al.}~\cite{ma2011coarse}. However, there is a quantitative difference between these two sets of results. Tian {\it et al.} found the energy released in the process of binding two nonprotonated (high pH) G4 PAMAM dendrimers to be $\sim$35 Kcal/mol and the equilibrium separation between the dendrimers to be 2.0 nm. In contrast, our simulation results predict a smaller interaction strength ($\sim$8 Kcal/mol) and a larger value of the equilibrium distance ($\sim$3.5 nm) for two G4 nonprotonated dendrimers. This difference may arise from differences between the  models considered in the studies. Monomer-resolved coarse-grained simulations carried out by Tian {\it et al.} are unlikely to capture the microscopic details of terminal group fluctuations and formation of the hydration layer near the dendrimer surface which play an important role in the effective interaction. In contrast, our full atomistic simulations do capture these microscopic details and are expected to provide more accurate results. 

\vspace{0.1cm}


We now consider the conformational changes in the dendrimer structure during the aggregation process. Instantaneous snapshots of the dendrimers when their centers of mass are coincident are shown in Fig. 3. We observe a strong overlap between the dendrimers which suggests that they act as soft flexible molecules, rather than hard colloidal objects. This observation is in agreement with previous coarse-grained simulation studies of interacting dendrimers ~\cite {huissmann2011effective}. When the dendrimers overlap, the branches of the dendrimers become intertwined with each other instead of simply interpenetrating. Because of strong steric interaction between the atoms, dendrimers open their branches (Fig. 3), so that the branches of one dendrimer can wrap around the branches of the other. To confirm the intertwining of the branches, we have calculated the number of close contacts between the dendrimers. The atoms of the first dendrimer which are within 3 {\AA} distance from any atom of the second dendrimer are considered to be in contact with the second dendrimer. The number of close contacts as a function of inter-dendrimer distance is shown in Fig. 4. We observe that when the centers of mass of the dendrimers coincide, the number of atoms in close contact is significantly smaller than the total number of atoms. Had the dendrimers completely penetrated each other, the number of close contacts would have been much larger.
 

\vspace{0.1cm}

In order to get a quantitative measure of this overlap, we have used an overlap function $O(r)$ defined in the same way as by  Hui\ss mann {\it et al.}~\cite {huissmann2011effective}. 
If an atom of a dendrimer crosses the perpendicular bisector plane of the line joining the centers of mass of the dendrimers, it is considered to be in the overlapping region. The corresponding overlap function, $O(r)$ is defined as two times the ratio between the number of overlapping atoms and the total number of atoms in a dendrimer. Fig. 5(a) shows plots of this overlap function for protonated dendrimers of different generation. For comparison, we have also shown plots of the overlap function of two homogeneous spheres of radius $R_g$ (the radius of gyration of a dendrimer), given by

\begin{equation} \begin {aligned} O(r) & =\frac {1}{16}\left(4+\frac {R}{R_g}\right)\left(2-\frac {R}{R_g}\right)^2 \hspace{10pt} R\le 2R_g \\ & = 0 \hspace {120pt} R\ge 2R_g\end {aligned}\end{equation} 

Clearly, the overlaps $O(r)$ between two dendrimers do not follow the overlap between two spheres of radius equal to the $R_g$ of a dendrimer, which suggests that the penetration of two aggregating dendrimers is not similar to that of two homogeneously charged spheres of radius equal to the $R_g$ of a dendrimer. This result contradicts the behavior of $O(r)$ observed by Hui\ss mann {\it et al.}~\cite {huissmann2011effective}. They found a very good similarity between the overlap function of the dendrimers and that of the spheres of radius equal to the $R_g$ of the dendrimer and hence, concluded that the interacting dendrimers can be thought of as homogeneously charged spheres in a coarse-grained description. They argued that the branches of the dendrimers do not retreat when the dendrimers approach each other, as the distributions of the monomers still remain homogeneous within the spheres around the centers of mass of the dendrimers. However, from our atomistic simulation, we observe that the atoms belonging to each dendrimer do retreat as the dendrimers approach each other because of the strong electrostatic and steric interactions between the atoms. 
To get a quantitative measure of this distortion, we have calculated the asphericity factor $\delta$ of the dendrimers using the following definition:

\begin{equation} \delta=1-3\frac{<I_2>}{<I_1^2>}\end{equation} \begin{equation} I_1=I_x+I_y+I_z\hspace {0.2cm}and\hspace {0.2cm} I_2=I_xI_y+I_xI_z+I_yI_z\end{equation}

where ($I_x$, $I_y$, $I_z$) are the eigenvalues of the shape tensor. The value $\delta=0$ and $\delta=1$ correspond to the cases where the atoms are in a spherical and in a linear configuration, respectively. 
 In Fig. 5(b), we observe that $\delta$ changes from a small initial value to a larger value as the distance between the dendrimers decreases, which clearly indicates that the dendrimers no longer remain spherical when they approach each other. Because of this deviation from spherical structures, the overlap function $O(r)$ between the dendrimers is not similar to the overlap between two homogeneous spheres.




\vspace{0.1cm}

In the previous section, we have discussed the structural changes of the dendrimers when they strongly interact with each other. Now, we investigate the functional form of the PMF between the dendrimers. The functional form of the effective interaction between two dendrimers is of particular interest because it may help in developing a coarse-grained model to study systems containing many dendrimers. Likos {\it et al.}~\cite{likos2002gaussian} proposed a theory for the effective interaction between dendrimers which is based on the monomer density profile of the dendrimers. Using a Flory-type argument, they derived a Gaussian effective interaction between a pair of G4 dendrimers given by 
\begin{equation}V_{eff}(R)=N^2v_0k_BT\left( \frac {3}{4\pi R^2_g}\right)^\frac {3}{2}\exp\left(-\frac{3R^2}{4R^2_g}\right)\end{equation}
where $N$, $v_0$, $R$ and $R_g$ are the number of monomers, excluded-volume parameter, inter-dendrimer distance and radius of gyration of a dendrimer, respectively. Subsequently, employing Monte-Carlo and molecular dynamics simulations, they observed ~\cite{gotze2004tunable} that the effective interaction between dendrimers can be fitted by a sum of two Gaussian functions,
\begin{equation}V_{eff}(R)=\epsilon_1 \exp\left(-\frac{3R^2}{4R^2_g}\right)+\epsilon_2 \exp\left[-\alpha (\frac{R}{R_g} - \gamma )^2 \right] \end{equation}
where $\epsilon_1$, $\epsilon_2$, $\alpha$ and $\gamma$ are  fitting parameters. We have fitted the PMF profiles of protonated dendrimers obtained from all atom simulation by both a Gaussian function and a sum of two Gaussian functions similar to equations 4 and 5, respectively. Fig. 6(a) clearly shows that the effective interaction between protonated dendrimer is not a Gaussian function. A sum of two Gaussian functions (equation 5) fits the PMF profile reasonably well (Fig. 6(b)) at relatively large center-to-center distances between the dendrimers. However, at strong overlap conditions (analogous to high density) when the dendrimers are very close to each other, a deviation between the PMF profile and the fitting function is observed.  Interestingly, a sum of an exponential function and a Gaussian function fits the PMF profile extremely well throughout the interaction region. The PMF profiles of the protonated dendrimers are fitted by the following equation:
\begin{equation}V_{eff}(R)=\epsilon_1 \exp(-\alpha_1\frac{R}{R_g})+\epsilon_2 \exp(-\alpha_2 \frac{R^2}{R_g^2})\end{equation}
Similarly, the PMF profiles of nonprotonated dendrimers were fitted by the following equation: 
\begin{equation}V_{eff}(R)=\epsilon_1 \exp(-\alpha_1\frac{R}{R_g})-\epsilon_2 \exp(-\alpha_2\frac{(R-R_0)^2}{R_g^2})\end{equation}
 
\begin{table}[t]
\centering 
    \begin{tabular}{lc@{\hspace{0.2cm}}c@{\hspace{0.2cm}}c@{\hspace{0.2cm}}c@{\hspace{0.2cm}}c@{\hspace{0.2cm}}c@{\hspace{0.2cm}}r}
    \hline
    Systems & $R_g (\AA)$ & $\epsilon_1$ & $\alpha_1$ & $\epsilon_2$ & $\alpha_2$ & $R_0$ \\ \hline
    G2P & 12.2 & 45.50 & 1.342 & 11.40 & 0.2679 & - \\ 
    G3P & 15.8 & 50.90 & 1.319 & 39.03 & 0.3495 & - \\ 
    G4P & 20.6 & 162.17 & 1.324 & 90.28 & 0.4668 & - \\ 
    G4P (0.1 M) & 20.0 & 87.04 & 1.161 & 60.79 & 0.6399 & - \\
    \hline
    G3NP & 12.3 & 276.29 & 1.805 & 22.91 & 0.4539 & 12.0 \\
    G4NP & 15.5 & 1103.9 & 1.659 & 62.42 & 0.4805 & 16.0 \\
    G4NP (0.1 M)& 15.4 & 1189.1 & 1.336 & 95.71 & 1.1384 & 24.52 \\   
    \hline
    \end{tabular}
\caption{\label{parameter} Values of the fitting parameters for equations 6 and 7}
\end{table} 
Fig. 7(a) and 7(b) show that these functional forms provide very good fits of the PMF data throughout the interaction region. The values of $R_g$ and the fitting parameters are shown in Table 1. Larger values of $\epsilon_1$ suggests that the exponential function dominates over the Gaussian function in the PMF profiles. This observation is contrary to the previous models~\cite{likos2002gaussian, gotze2004tunable} which predicted the nature of the effective interaction to be Gaussian. These models are based on the monomer density profile of a single dendrimer and are valid at low concentrations where the density profiles of interacting dendrimers remain nearly the same as the density distribution of a non-interacting dendrimer. However, at high concentrations, the dendrimers interact strongly with each other  and their density distribution changes. In the previous section, we have discussed the configurational changes that occur in the dendrimer structures as they interact with each other. We would also like to point out that the solvent may play an important role in the effective interaction. Thus, a realistic model of interacting dendrimers should include the effects of charges, solvent and changes in the structures (density distribution) as they approach each other.

\vspace{0.1cm}


To investigate the effect of salt concentration on the effective interaction between the dendrimers, we have calculated the PMF between two G4 dendrimers at 100 mM salt concentration. Protonated dendrimers have net positive charge. So the electrostatic interaction contributes  significantly to the effective interaction between protonated dendrimers. So at high salt concentration, because of the screening effect, the strength of the repulsive interaction between the dendrimers diminishes which can be observed in Fig. 8(b). To estimate the contribution of the electrostatic interaction in the effective interaction between the protonated dendrimers, we have fitted the protonated dendrimer PMF profile using the following equation:

 \begin{equation} PMF(100 \hspace {2mm}mM) = a_0 \times {PMF(0 \hspace {2mm}mM)} + (1-a_0) \times {PMF(0 \hspace {2mm}mM)} \times {exp(-KR)} \end{equation}

where $PMF(100 \hspace {2mm}mM)$ and $PMF(0 \hspace {2mm}mM)$ represent the PMF between protonated dendrimers at 100 mM and 0 mM salt concentration, respectively. $K^{-1}$ and $a_0$ are the Debye screening length and fitting parameter, respectively. We use $K$=0.104 corresponding to a Debye length of 9.6 $\AA$ at 100 mM salt concentration and 300 K temperature. The first term in the equation 8 represents the non-electrostatic component which we assume to be unchanged at higher salt concentration. And the second term represents the electrostatic component which deceases exponentially at higher salt concentration due to screening. The best fitting (inset of figure 8(b)) gives the value of the $a_0$ parameter to be 0.46. Thus we conclude approximately $54\%$ contribution to the protonated dendrimer PMF comes from the electrostatic interaction between the dendrimers. Note that the fitting is reasonably well up to 10 $\AA$ center-to-center distance. Beyond that distance, the fitting line deviates due to the fact that we have assumed non-electrostatic component does not change at higher salt concentration. However, non-electrostatic component also may change due to the structural change of the dendrimers at higher salt concentration. \\

In contrast, van der Waals interactions mainly contribute to the interaction between nonprotonated dendrimers. Thus we observe that the interaction strength at higher salt concentration remains almost same up to 25 $\AA$ center-to-center distance (figure 8(a)). Beyond that distance, interaction strength increases which could be due to structural change of the dendrimers at high salt concentration. Note that the interaction strength at the attractive region remains almost same (inset of figure 8(a)). This is due to the fact that in this region, van der Waals interaction is dominating force as we discussed before. Thus salt concentration has negligible effect on the attractive strength between nonprotonated dendrimers. As before, we have fitted the PMF by equations 6 and 7. Fitting parameters are given in Table 1.

\subsection{CONCLUSIONS}
In summary, employing fully atomistic MD simulations, we have calculated the effective interaction between two PAMAM dendrimers. The PMF between the dendrimers  depends strongly on the protonation level and the size of the dendrimer. There is a global minimum in the PMF profile of nonprotonated dendrimers which represents the attractive nature of the effective interaction for these dendrimers. We argue that the origin of this attractive region is due to the van der Waals interaction between the nonprotonated dendrimers. On the other hand, the effective interaction between protonated dendrimers are repulsive throughout the interaction region. Due to the net positive charges of protonated dendrimers, the electrostatic force between the protonated dendrimers is strong which makes these dendrimers to be repulsive even at a region where dendrimers do not overlap. It would be interesting to decrease the protonation level of the dendrimer gradually and to find out the critical value of the protonation level at which the attraction between the dendrimer arises. We plan to take up this work in our future study. The PMF profiles of the dendrimers are fitted very well by a sum of an exponential and a Gaussian function, with the strength of the exponential function much larger than that of the Gaussian function. Earlier studies using simplified models predicted the effective interaction to be Gaussian. However, we observe that charges residing on the dendrimers and atomic-scale fluctuations of the local density, which were not included in previous models, significantly contribute to the PMF and make its profile non-Gaussian. Our fully atomistic simulations provide important information regarding the strength and nature of the effective interaction between two dendrimers. We expect that this information will help towards developing a coarse-grained model to investigate the collective properties of systems of many dendrimers.

\vspace{0.1cm}
We acknowledge financial support from DST, India. T.M. thanks Council of Scientific and Industrial Research (CSIR), India for fellowship.


\begin{thebibliography}{10}
\bibitem{footnote1}%
  \BibitemOpen
  \bibinfo {journal} {To compute the van der Waals energy contribution, we set
  the charge of each atom to be zero. First we generate the trajectories using
  the original potential. Then using these atom positions, we calculated the
  van der Waals energy by setting the atomic charges and bonded interaction
  parameters to be zero.}\BibitemShut{Stop}%
\end{thebibliography}

%

\newpage

\begin{figure}
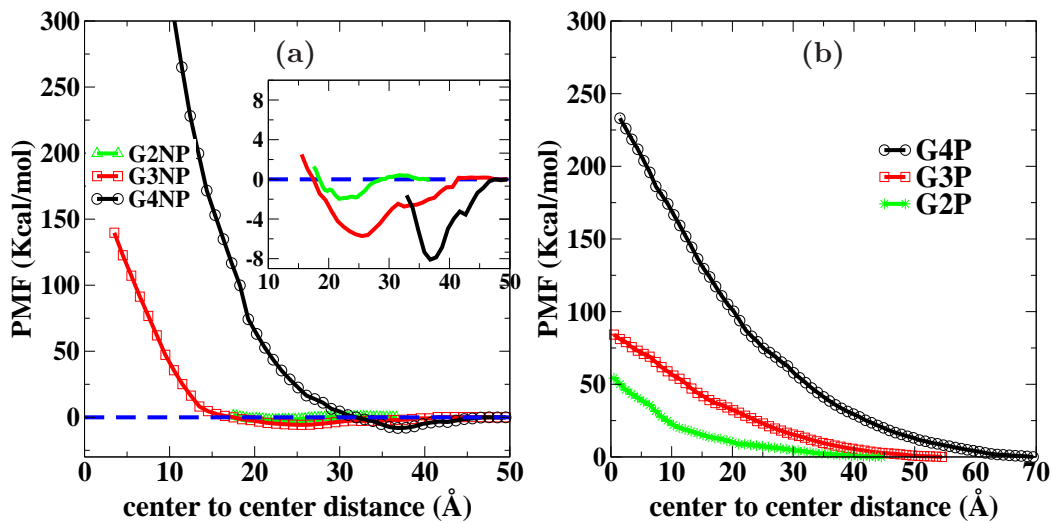

\begin{center} 
\begin{tabular}{cc}\includegraphics[height=2.7in, width=2.7in]{nonprot.eps} \put(-95,175){\textbf{(a)}}
 & \includegraphics[height=2.7in, width=2.7in]{prot.eps} \put(-95,175){\textbf{(b)}}
\end{tabular} \end{center} 
\caption {PMF between two dendrimers as a function of the center-to-center distance for (a) nonprotonated and (b) protonated case. Inset of (a) shows the attractive region of the PMF for two nonprotonated dendrimers.} 
\end {figure} 

\begin{figure}
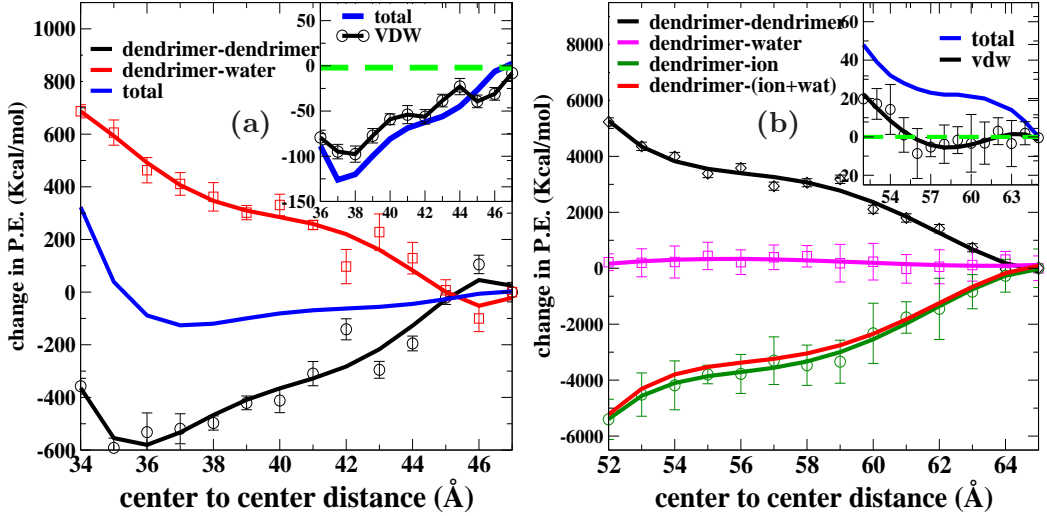

\begin{center} 
\begin{tabular}{cc}\includegraphics[height=2.7in, width=2.7in]{npsplit.eps} \put(-110,145){\textbf{(a)}}
 & \includegraphics[height=2.7in, width=2.7in]{psplit.eps} \put(-110,145){\textbf{(b)}}
\end{tabular} \end{center} 
\caption {(a) Potential energy contributions of the various components in the attractive region of the G4 nonprotonated dendrimer PMF profile. Square and circles are the data points. Solid lines are the polynomial fitting. Black and red lines show the change in dendrimer-dendrimer and
dendrimer-water potential energy, respectively. Inset shows the contribution of the van der Waals interactions in the total potential energy change. (b) Various contributions to the potential energy of interaction between the G4 protonated dendrimers. Diamond, square and circles are the data points. Solid lines are the polynomial fitting. Black, magenta and green solid lines are the change in the potential energy of dendrimer-dendrimer, dendrimer-water and dendrimer-ion interactions, respectively. Solid red line shows the total change in dendrimer-water and dendrimer-ion potential energy. Inset shows the contribution of the van der Waals interactions to the total potential energy change.} 
\end {figure}

\begin{figure}\begin{center} \begin{tabular}{cc}\includegraphics[height=2.7in, width=2.7in]{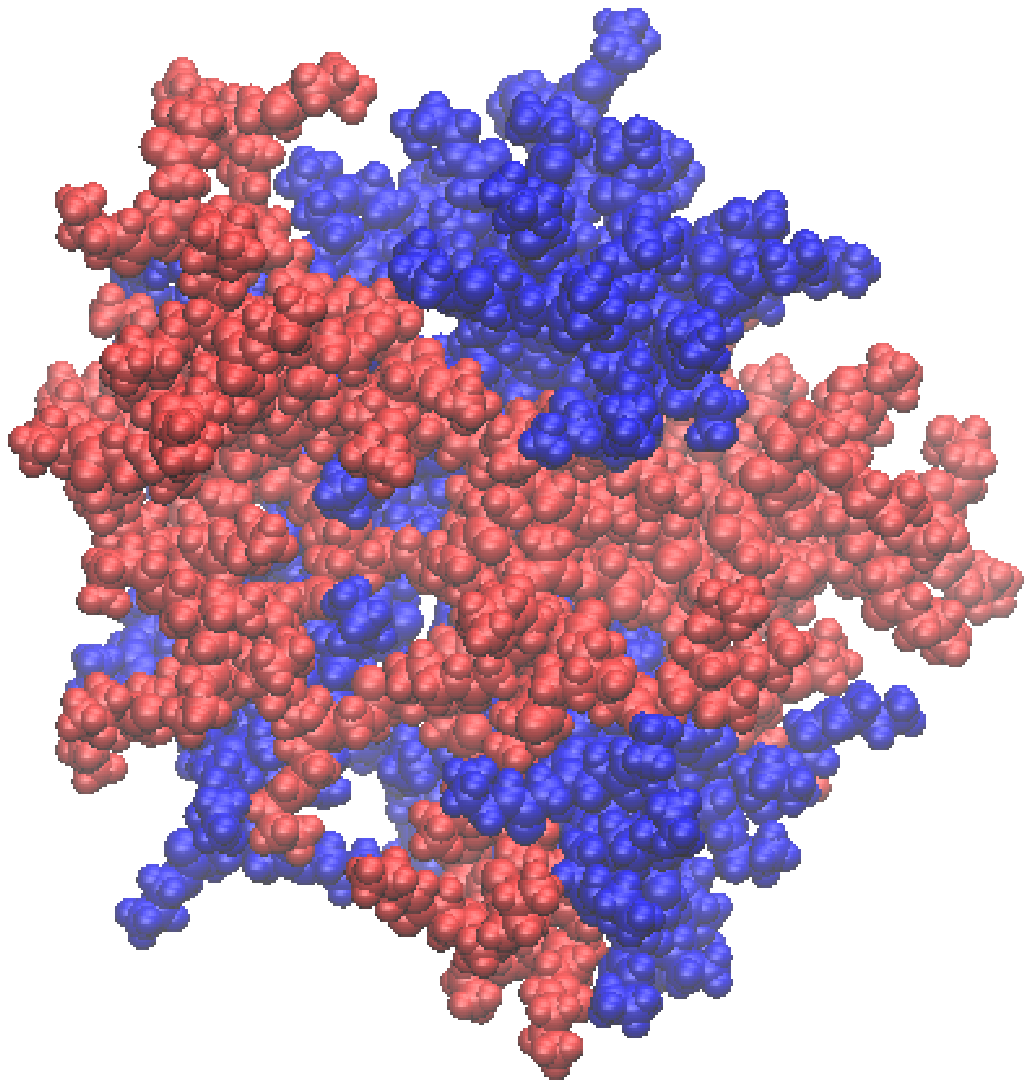} & \includegraphics[height=2.7in, width=2.7in]{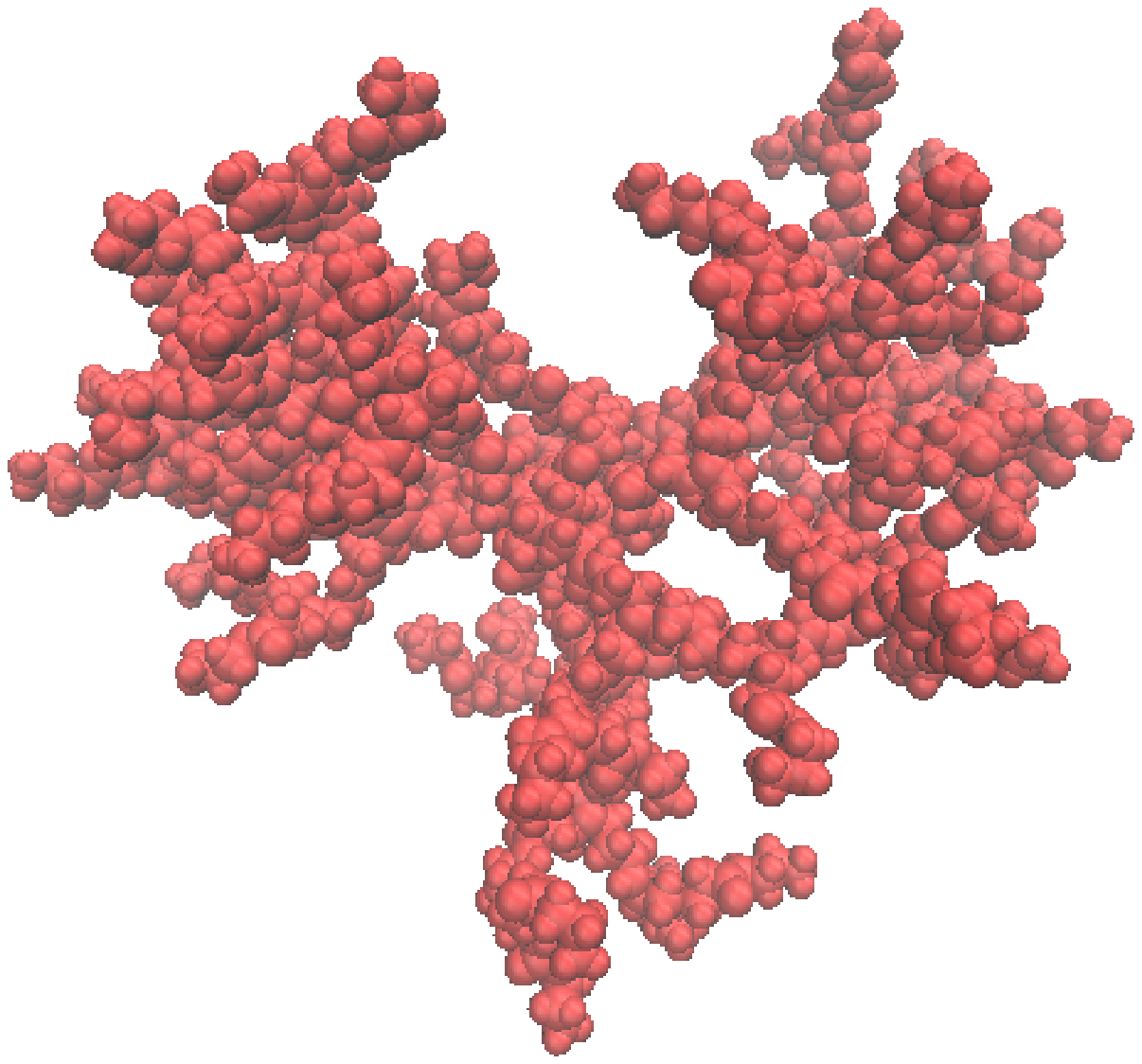} \end{tabular} \end{center} \caption {Instantenous snapshot of the interacting G4 protonated dendrimers at 0 {\AA} seperation. Opennig of the branches is shown in the right panel.} \end {figure} 

\begin{figure}\begin{center} \includegraphics[height=3.0in, width=3.0in]{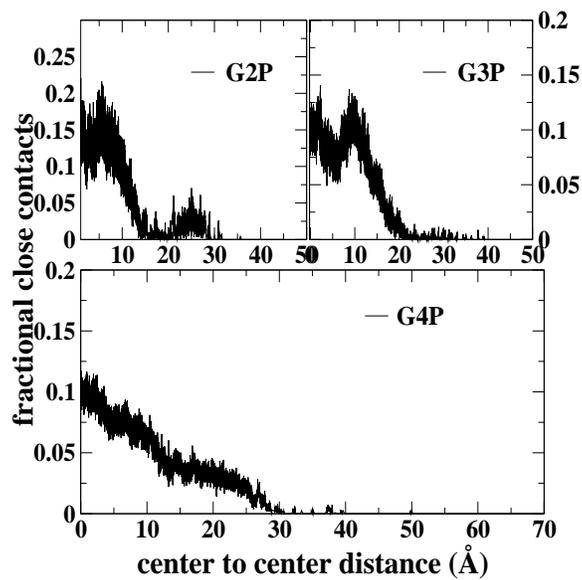} \end{center} \caption {Number of close contacts between the protonated dendrimers.} \end {figure} 

\begin{figure}
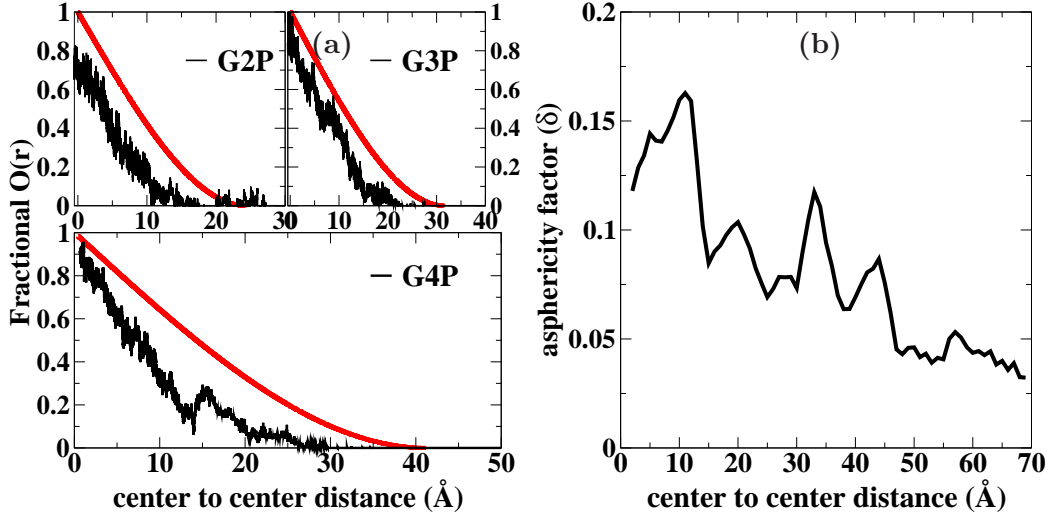

\begin{center} \begin{tabular}{cc} \includegraphics[height=2.7in, width=2.7in]{overlap.eps} \put(-80,175){\textbf{(a)}}
 & \includegraphics[height=2.7in, width=2.7in]{sphericityg4p.eps} \put(-95,175){\textbf{(b)}}
 \end{tabular}\end{center} 
\caption {(a) Overlap function $O(r)$ for different dendrimers. (b) Asphericity factor ($\delta$) of a dendrimer as a function of inter-dendrimer distance. Red lines in (a) show  $O(r)$ for two homogeneous spheres of radius equal to the $R_g$ of the dendrimers.} \end {figure}

\begin{figure}
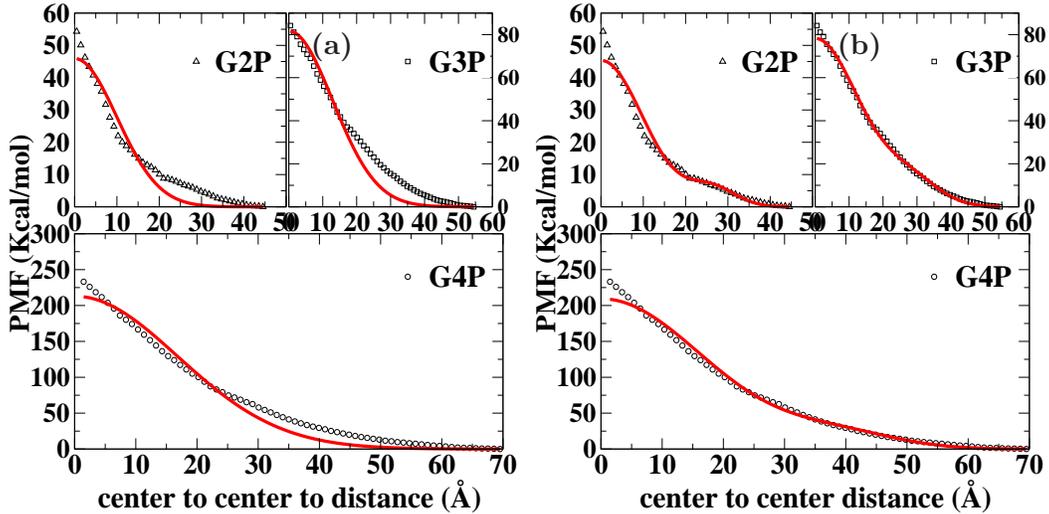
\begin{center} \begin{tabular}{cc} \includegraphics[height=2.7in, width=2.7in]{prot1.eps} \put(-80,175){\textbf{(a)}}
& \includegraphics[height=2.7in, width=2.7in]{prot2.eps} \put(-80,175){\textbf{(b)}}
\end{tabular}\end{center} 
\caption {PMF profiles of protonated dendrimers are fitted by (a) a Gaussian function (equation 4) and (b) a sum of two Gaussian functions (equation 5)}\end {figure}

\begin{figure}
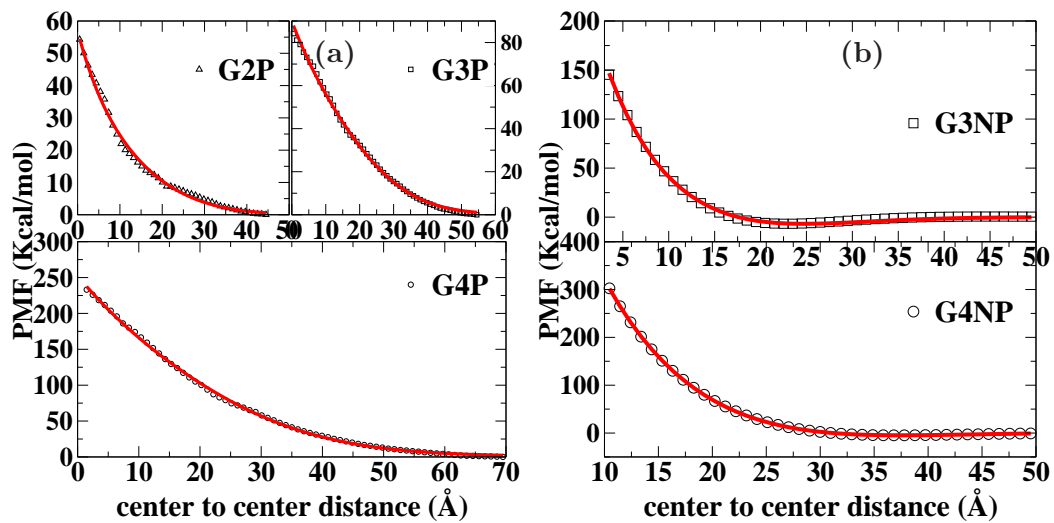
\begin{center} \begin{tabular}{cc} \includegraphics[height=2.7in, width=2.7in]{prot3.eps} \put(-80,175){\textbf{(a)}}
 & \includegraphics[height=2.7in, width=2.7in]{nonprot1.eps} \put(-80,175){\textbf {(b)}}
 \end{tabular}\end{center} 
\caption {PMF profiles of dendrimers are fitted by a sum of exponential and Gaussian functions, equation 6 and 7 for (a) protonated dendrimers and (b) nonprotonated dendrimers}\end {figure}

\begin{figure}
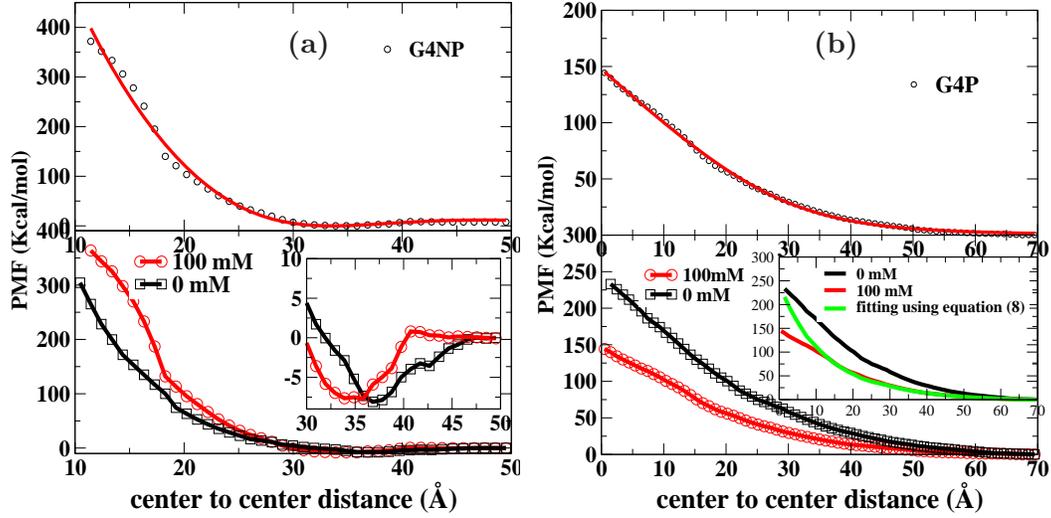
 \begin{tabular}{cc}\includegraphics[height=2.7in, width=2.7in]{g4npsalt.eps} \put(-90,175){\textbf{(a)}}
 & \includegraphics[height=2.7in, width=2.7in]{g4psalt.eps} \put(-90,175){\textbf{(b)}}
\end{tabular}  \caption {Effect of salt concentration on the PMF between (a) nonprotonated and (b) protonated G4 dendrimer. Inset of (a) shows the attractive region. Strength of the interaction in this part remains almost same for both the 0 mM and 100 mM cases, which suggests that the contribution from the electrostatic interaction in this region is negligible for the nonprotonated case. The strength of the repulsive interaction between protonated dendrimers decreases compared to 0 mM case due to screening of the electrostatic interaction at 100 mM concentration. Inset of (b) shows the fitting of 100 mM PMF using equation (8) (see text). Upper panels of (a) and (b) show the fitting of the PMF at 100 mM salt concentration using equation 6 (for protonated dendrimer) and 7 (for nonprotonated dendrimer).} \end {figure}

\end{document}